\documentclass[a4paper,prl,reprint,superscriptaddress]{revtex4-1}
\usepackage{ulem}
\usepackage{amssymb}
\usepackage{amsmath}
\usepackage{epsfig}
\usepackage{color}
\usepackage{graphics, graphicx}
\usepackage{bbold}
\usepackage{psfrag}
\usepackage{mathcomp}
\usepackage{amsmath}
\usepackage{amssymb}%
\usepackage{mathrsfs}%
\usepackage{subfigure}
\usepackage{verbatim}
\usepackage{xcolor}
\usepackage[colorlinks,citecolor=blue,urlcolor=blue]{hyperref}
\usepackage{mathrsfs}
\usepackage[margin=1in,letterpaper]{geometry} % decreases margins
\usepackage{ulem}

\begin{document}

%	\title{\textbf{Hyperbolic fringes in quasiparticle interference patterns with two impurities in superconductors}}
	\title{\textbf{Hyperbolic fringe signal for twin impurity quasiparticle interference}}
	\author{Peize Ding}
	\email{pd2714@columbia.edu}
	\affiliation{Institute for Theoretical Physics,
		University of W\"{u}rzburg, Am Hubland, D-97074 W\"{u}rzburg, Germany}
	\affiliation{School of the Gifted Young, University of Science and Technology of China, Hefei 230026, China}
	\affiliation{Department of Physics, Columbia University, New York 10027, NY, USA}
	\author{Tilman Schwemmer}
	\affiliation{Institute for Theoretical Physics,	University of W\"{u}rzburg, Am Hubland, D-97074 W\"{u}rzburg, Germany}
	\author{Ching Hua Lee}
	\affiliation{Department of Physics, National University of Singapore, Singapore, 117542}
	\affiliation{Joint School of National University of Singapore and Tianjin University, International Campus of Tianjin University, Binhai New City, Fuzhou 350207, China}
	\author{Xianxin Wu}
	\email{xxwu@itp.ac.cn}
	\affiliation{CAS Key Laboratory of Theoretical Physics, Institute of Theoretical Physics, Chinese Academy of Sciences, Beijing 100190, China}
    %\affiliation{Max-Planck-Institut f\"ur Festk\"orperforschung, Heisenbergstrasse 1, D-70569 Stuttgart, Germany}

	\author{Ronny Thomale}
	\email{rthomale@physik.uni-wuerzburg.de}
	\affiliation{Institute for Theoretical Physics,
		University of W\"{u}rzburg, Am Hubland, D-97074 W\"{u}rzburg, Germany}
	
	\date{\today}
	
%	\begin{abstract}
%		We study the quasiparticle interference pattern emanating from two impurities on the surface of superconductors. Through both numerical and analytical investigation, we find that hyperbolic fringes, with the location of the two impurities being the focus points, appear in the local density of states distribution for (i) conventionally pairing superconductors with magnetic impurities, (ii) $ p+ip $, $ d+id $-wave chiral superconductors with non-magnetic (or magnetic) impurities, and (iii) extended $s$-wave pairing iron-based superconductors around non-magnetic (or magnetic) impurities with sign change on different Fermi pockets. We demonstrate that the emergence of hyperbolic fringes (around two non-magnetic impurities) in gapped single-pocket system undoubtedly indicates time-reversal symmetry breaking in the gap function. In gapped multi-pocket systems, the appearance of hyperbolic fringes is also helpful to detect sign change in the superconducting order parameters.
%	\end{abstract}
	
	\begin{abstract}
		We study the quasiparticle interference (QPI) pattern emanating from a pair of adjacent impurities on the surface of a gapped superconductor (SC). We find that hyperbolic fringes (HF) in the QPI signal can appear due to the loop contribution of the two-impurity scattering, where the location of the two impurities are the hyperbolic focus points. For a single pocket Fermiology, an HF pattern signals chiral SC order for non-magnetic impurities and requires magnetic impurities for non-chiral SC. For a multi-pocket scenario, a sign-changing order parameter such as $s_{\pm}$-wave likewise yields an HF signature. We discuss twin impurity QPI as a new tool to complement the analysis of superconducting order from local spectroscopy.
	\end{abstract}

	\maketitle
	
	\twocolumngrid
	
	\textit{Introduction.---} Quasiparticle interference (QPI) around impurities, which can be probed through a scanning tunneling microscopy (STM) measurement, has acquired a pivotal role in exploring the properties of unconventional electronic states of matter including high-$ T_c $ superconductors (SCs)~\cite{extrema, exp_cup, Dunghai, iron_expm, Sykora, Weyl, PhysRevB.82.224506, BENA2016302} and topological insulators (TIs)~\cite{exp_ti, Shoucheng}. For SC, the local density of states (LDOS) modulation patterns not only reflect principal features of electronic dispersion and pairing symmetries~\cite{extrema, RevModPhys.78.373, bound_yu, bound_dwave}, but are also sensitive to sign changes in the SC order parameter~\cite{Dunghai, iron_expm, Sykora} as well as to time-reversal symmetry~\cite{Shoucheng, PhysRevLett.121.176401}. 
	%Especially, in ref.~\cite{Aharonov}, the authors studied the LDOS distribution around two impurities on a two-dimensional metallic sample in perpendicular magnetic field. They found that Aharonov-Bohm type of oscillations, which only depend on the magnetic flux penetrating through the triangle formed by the two impurities and the STM tip, appear in the LDOS distribution. The pattern they found is closely related to TRS breaking effect caused by magnetic flux. Therefore, it is natural to ask the question, what LDOS modulation patterns would appear for chiral superconductors with the presence of two impurities, which breaks TRS without magnetic field. Also, would the QPI patterns around two impurities reflect other properties like sign change in superconductors?
	%However, the rich physics contained in the QPI patterns around two impurities in various kinds SC has not yet been thoroughly investigated.
	In addition to QPI patterns arising from a single impurity, the LDOS distribution and possible bound states due to the presence of multiple impurities have been studied in a variety of systems \cite{two_s1, two_s2, two_p, two_dwave, Aharonov, PhysRevB.105.245403}.

	The QPI pattern's phase sensitivity renders it preeminently suited to resolve intricate properties of an unconventional SC state of matter. At present, the evidence on the nature of unconventional pairing symmetry of several material classes is still incomplete. This includes candidates for chiral SC order such as strontium ruthenate~\cite{pusto}, Na-doped cobaltates~\cite{PhysRevLett.111.097001} or, more lately, Sn/Si heterostructures~\cite{PhysRevLett.128.167002} and kagome metals~\cite{kagomereview}. Likewise, additional ways to track the sign-changing nature of extended s-wave order, as suspected for many iron pnictide families, are highly sought after~\cite{RevModPhys.83.1589}.

	In this Letter, we propose a hyperbolic fringe (HF) signal fingerprint found in the QPI pattern of two adjacent impurities deposited on a gapped SC, which we coin {\it twin QPI}. We find that through the HF signal, the twin QPI pattern allows to retrieve information beyond the single impurity case, and thus to draw conclusions on the either chiral or multi-pocket sign changing nature of the SC order parameter.

    %	we provide a detailed investigation of the QPI patterns in conventional, chiral, and iron-based SC, with the presence of two non-magnetic or magnetic impurities. We find that a special feature---hyperbolic fringes (HF)---emerges for (i) $ s $-wave pairing state with magnetic impurities, (ii) chiral $ p $ and $ d $-wave pairing SC with non-magnetic (or magnetic) impurities, and (iii) iron-based SC with sign change in order parameter with non-magnetic (or magnetic) impurities. When the Fermi surface (FS) is nearly circular, we derive analytical formulas to explain the patterns. We emphasize that our finding opens up a new possibility to utilize QPI pattern around two impurities to detect TRS breaking in superconducting order parameters in gapped single-pocket systems, which is beyond the capability of QPI around single impurity. The emergence of the HF also provides an exotic way to detect sign change in gapped multi-pocket systems. Finally, we discuss the generality of our results through considering the relevance of the HF in other systems, as well as possibilities to detect this signal experimentally.

%\begin{widetext}
		\begin{figure*}[t]
		\centering
		\includegraphics[width=0.7\textwidth]{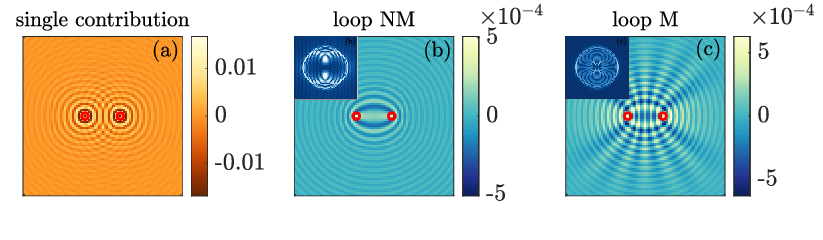}
    	\caption{
    	Twin impurity QPI for a single-pocket $s$-wave SC ($\mu = -3$,  $\Delta = 0.1$, $\omega = 1.1\Delta$, $d=18$) in real space. Note that the shown LDOS is measured in units of $(t a^2)^{-1}$ and thus dimensionless for our choice of $a=1$ and $t=1$.
    	%(a) First Brillouin zone plot of the band structure and the FS (denoted by red line) nearby the band edge. 
    	(a) Single impurity contribution to the QPI pattern. (b) Loop contribution to QPI for non-magnetic (NM) twin impurities. (c) Loop contribution to QPI for magnetic (M) twin impurities. The insets show the Fourier transformation of the patterns. A hyperbolic fringe (HF) pattern emerges on top of the elliptical fringe background.}
		\label{s-wave}
	\end{figure*}
	%\end{widetext}
 
\begin{figure*}[t]
		\centering
		\includegraphics[width=1\textwidth]{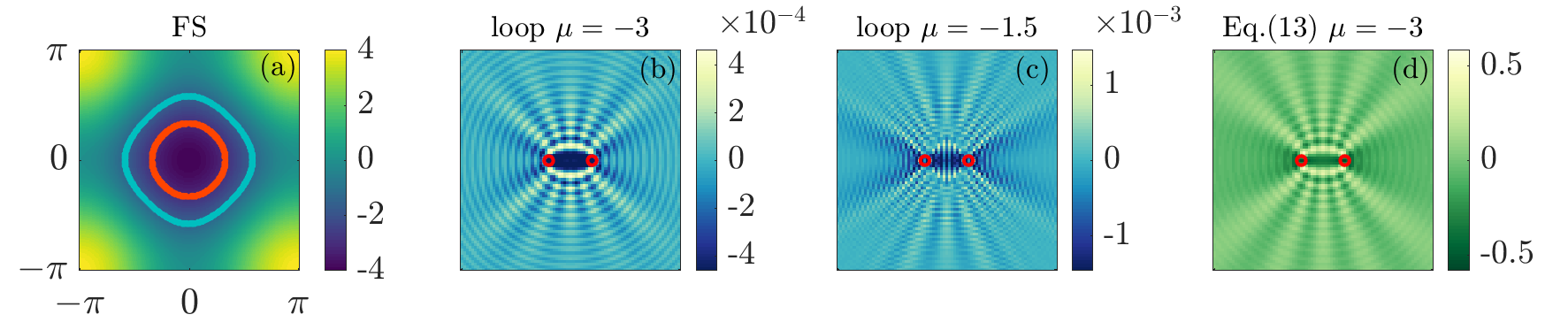}
		\caption{Twin impurity QPI for a single-pocket chiral SC ($\Delta = 0.05$, $\omega = 1.1\Delta$, $d = 16$). (a) Band dispersion $\epsilon(\mathbf{k}) = -2 (\cos k_x + \cos k_y)$ (denoted by color) and the FS at $\mu = -3$ (red line) and  $\mu = -1.5$ (blue line) in the first Brillouin zone.
		(b), (c) Twin non-magnetic impurity loop contribution for $ p+ip $-wave pairing with $\mu=-3$ and $\mu = -1.5$. (d) A plot of our analytical expression (Eq.~(\ref{simulation_p})) for the loop contribution (in arbitrary unit) on a $ 81 \times 81 $ real space lattice with $\mu = -3$ shows the expected agreement with the corresponding numerical results presented in panel (b).}
		\label{loop_p}
	\end{figure*}

	\textit{Minimal model.---}
	In order to complement the numerical analysis with an analytically tractable limit to showcase the mathematical structure of the HF pattern, we initially constrain ourselves to the simplest Bardeen-Cooper-Schrieffer (BCS) Hamiltonian for a single electronic band with nearest neighbor hybridization on a square lattice $ \varepsilon_{\mathbf{k}} = - 2 t (\cos k_x + \cos k_y) - \mu $:
	\begin{equation}\label{Hamiltonian}
	H = \sum_{\mathbf{k}s}\varepsilon_{\mathbf{k}}^{} c_{\mathbf{k}s}^{\dagger}c_{\mathbf{k}s}^{} + \sum_{\mathbf{k}}(\Delta_{\mathbf{k}}c^{\dagger}_{\mathbf{k}\uparrow}c^{\dagger}_{-\mathbf{k}\downarrow} + h.c.),
	\end{equation}
    where we set the hopping parameter $t=1$ and the lattice spacing $a=1$.
	%We assume the two spin species to be degenerate, hence the $2\times2$ Nambu space. 
	Initially, we assume the chemical potential to be located at fillings where the Fermi surface is approximately circular (Fig.~\ref{loop_p} (a) red line). 
	The presence of impurity {scattering} is modeled by
	\begin{equation}\label{key}
	H^{\text{imp}}_{\text{NM/M}} = U_{0}\sum_{\mathbf{r}}\sum_{n}\delta_{\mathbf{r}, \mathbf{r}_{n}} (c^{\dagger}_{\mathbf{r}\uparrow}c^{}_{\mathbf{r}\uparrow} \pm
	c^{\dagger}_{\mathbf{r}\downarrow}c^{}_{\mathbf{r}\downarrow}),
	\end{equation}
	where $ \mathbf{r}_{n} $ denotes the location of impurities, $U_0$ is the impurity coupling strength, and {the $ +/- $ sign corresponds to non-magnetic/magnetic impurities.} 
	%$ \pm $ corresponds to non-magnetic (NM) and magnetic (M) impurities, respectively.
	The QPI pattern observed by an STM measurement is related to the LDOS distribution $ N(\mathbf{r}, \omega) $, where we recast $\omega$ as the frequency bias of the STM tip. It reads
	\begin{equation}\label{LDOS}
	N(\mathbf{r}, \omega) = -\frac{1}{\pi}\textrm{Im Tr} \left[\frac{\hat{1} + \hat{\tau}_{z}}{2}\mathbf{G}(\mathbf{r}, \mathbf{r}, \omega + i\eta)\right],
	\end{equation} %\frac{\mathbb{1} + \sigma_z}{2}
    where $ \mathbf{G} $ denotes the Nambu Green's function, with the four-component spinor operator given by $\left(c_{\mathbf{k}, \uparrow}^{\dagger} c_{-\mathbf{k}, \downarrow}^{\dagger} c_{-\mathbf{k}, \downarrow}-c_{\mathbf{k}, \uparrow}\right)^{\dagger}$ in momentum space. ${\vec{\tau}}=(\tau_x, \tau_y, \tau_z)$ denotes the Pauli matrix vector in particle-hole space, and $ \eta $ denotes an infinitesimal positive number regulator. In the absence of impurities, the Green's function is given by
	%\begin{equation}\label{Green}
	%\mathbf{G}_{{0}}\left(\mathbf{r}, \omega\right)\equiv
	%\frac{1}{\mathcal{S}} \sum_{\mathbf{k}} \mathrm{e}^{i \mathbf{k} \cdot\mathbf{r}}
	%\left[\omega \hat{1}-\varepsilon_{\mathbf{k}} \hat{\tau}_{3}-\Delta_{\mathbf{k}} %\hat{\tau}_{1}\right]^{-1}
	%\end{equation}
	\begin{equation}\label{Green}
	\begin{split}
	&\mathbf{G}_{{0}}\left(\mathbf{r}, \omega\right)\equiv\\
	&\left(\begin{array}{cccc}
    G_{0}\left(\mathbf{r}, \omega\right) & 0 & F_{0}\left(\mathbf{r}, \omega\right) & 0 \\
    0 & G_{0}\left(\mathbf{r}, \omega\right) & 0 & F_{0}\left(\mathbf{r}, \omega\right)\\
    \tilde{F}_{0}\left(\mathbf{r}, \omega\right) & 0 & -G_{0}\left(\mathbf{r}, -\omega\right) & 0\\
    0 & \tilde{F}_{0}\left(\mathbf{r}, \omega\right) & 0 & -G_{0}\left(\mathbf{r}, -\omega\right)
    \end{array}\right),
	\end{split}
	\end{equation}
	where
	\begin{equation}
	    \left(\begin{array}{c}
        G_{0}\left(\mathbf{r}, \omega\right) \\
        F_{0}\left(\mathbf{r}, \omega\right) \\
        \tilde{F}_{0}\left(\mathbf{r}, \omega\right)
        \end{array}\right)=\frac{1}{\mathcal{S}} \sum_{\mathbf{k}} \frac{\mathrm{e}^{i \mathbf{k} \cdot\mathbf{r}}}{\omega^{2}-\varepsilon_{\mathbf{k}}^{2}-\Delta_{\mathbf{k}}^{2}+i {0^{+}}}\left(\begin{array}{c}
        \omega+\varepsilon_{\mathbf{k}} \\
        \Delta_{\mathbf{k}} \\
        \Delta^{*}_{\mathbf{k}}
        \end{array}\right),
	\end{equation}
	and $ \mathcal{S} $ is the area of the sample. In the presence of impurities, the full Green's function is expanded with respect to $ U_{0} $ to obtain the infinite series
	\begin{equation}\label{key}
	\begin{aligned}
	\mathbf{G}(\mathbf{r}, \mathbf{r}^{\prime}) =& \mathbf{G}_{0}(\mathbf{r}- \mathbf{r}^{\prime}) + \sum_{\mathbf{r}^{\prime \prime}} \mathbf{G}_{0}\left(\mathbf{r}-\mathbf{r}^{\prime \prime}\right) U\left(\mathbf{r}^{\prime \prime}\right) \mathbf{G}_{0}\left(\mathbf{r}^{\prime \prime}-\mathbf{r}^{\prime}\right) \\
	&+\sum_{\mathbf{r}^{\prime \prime}\mathbf{r}^{\prime \prime \prime}} \mathbf{G}_{0}\left(\mathbf{r}-\mathbf{r}^{\prime \prime}\right) U\left(\mathbf{r}^{\prime \prime}\right) \\
	& \times \mathbf{G}_{0}\left(\mathbf{r}^{\prime \prime}-\mathbf{r}^{\prime \prime \prime}\right) U\left(\mathbf{r}^{\prime \prime \prime}\right) \mathbf{G}_{0}\left(\mathbf{r}^{\prime \prime \prime}-\mathbf{r}^{\prime}\right)+\ldots,
	\end{aligned}
	\end{equation}
	where $ U(\mathbf{r}) = U\sum_{n}\delta_{\mathbf{r}, \mathbf{r}_{n}} $ with $U=U_0\tau_z$ for non-magnetic and $U=U_0\sigma_z$ ($\vec{\sigma}=(\sigma_x,\sigma_y,\sigma_z)$ is the Pauli matrix vector in spin space) for magnetic impurities. { Note that for scattering around a pair of magnetic impurities, unless otherwise stated, we assume that their magnetic moments are aligned.
}
	%\begin{equation}\label{key}
	%U = 
	%\left\{
	%\begin{array}{lr}
	%U_{0}\sigma_z, & \textrm{non-magnetic}  \\
	%U_{0}I, & \textrm{magnetic}
	%\end{array}
	%\right.
	%\end{equation}
	%and 

	{\it Hyperbolic fringes in twin QPI.---}
	Assume two impurities to be located at $ \mathbf{r}_1 $ and $ \mathbf{r}_2 $. While there are infinitely many terms, we can separate them into two classes. The first contains all processes from $ \mathbf{r} $ to $ \mathbf{r}_1 $ ($ \mathbf{r}_2 $), and after possible multi-bouncing, back to $ \mathbf{r} $ from $ \mathbf{r}_1 $ ($ \mathbf{r}_2 $). We call this class \textit{single contribution}. The other part contains processes from $ \mathbf{r} $ to $ \mathbf{r}_1 $ ($ \mathbf{r}_2 $), and after possible multi-bouncing, back to $ \mathbf{r} $ from $ \mathbf{r}_2 $ ($ \mathbf{r}_1 $), which we call \textit{loop contribution}. In each class, the terms are summed up as a geometric series. The multi-bouncing processes at one impurity only renormalize $U$ according to $U'=[\hat{1}-U\mathbf{G}_0(0)]^{-1}U $.
	Summing up all the multi-bouncing processes between two impurities yields a second step renormalization of $U'$ according to
	$U^{''}=[\hat{1}-U^{'}\mathbf{G}_0(\mathbf{r}_1-\mathbf{r}_2)U^{'}\mathbf{G}_0(\mathbf{r}_2-\mathbf{r}_1)]^{-1}U^{'}
	$.
%	Finally, 
	We get
%	\begin{equation}\label{full}
%	\begin{split}
%	&\mathbf{G}(\mathbf{r}, \mathbf{r})-\mathbf{G}_0(0)=\\	&\mathbf{G}_0(\mathbf{r}-\mathbf{r}_1)U^{''}\mathbf{G}_0(\mathbf{r}_1-\mathbf{r})+(1\leftrightarrow 2)+ \\
%	&\mathbf{G}_0(\mathbf{r}-\mathbf{r}_2)U^{'}\mathbf{G}_0(\mathbf{r}_2-\mathbf{r}_1)U^{''}\mathbf{G}_0(\mathbf{r}_1-\mathbf{r})+(1\leftrightarrow 2)
%	\end{split}
%	\end{equation}
	\begin{eqnarray}
&&\mathbf{G}(\mathbf{r}, \mathbf{r})-\mathbf{G}_0(0)= \nonumber \\	&&\mathbf{G}_0(\mathbf{r}-\mathbf{r}_1)U^{''}\mathbf{G}_0(\mathbf{r}_1-\mathbf{r})+(1\leftrightarrow 2)+ \label{single}\\
	&&\mathbf{G}_0(\mathbf{r}-\mathbf{r}_2)U^{'}\mathbf{G}_0(\mathbf{r}_2-\mathbf{r}_1)U^{''}\mathbf{G}_0(\mathbf{r}_1-\mathbf{r})+(1\leftrightarrow 2),\label{loop}
	\end{eqnarray}
	where the lines~(\ref{single}) and~(\ref{loop}) denote {single} and {loop contribution}, respectively. 	We assume a gapped single-pocket $s$-wave SC $\Delta_{\mathbf{k}}^{s}=\Delta$ and an electronic filling close to the band edge such that we obtain an approximately circular Fermi surface (Fig.~\ref{loop_p}(a) red line). This allows us to gain an analytical grasp on $ \mathbf{G}_{{0}}\left(\mathbf{r}, \omega\right) $. Setting $ k_F = 1 $, we find~\cite{supp}
	\begin{equation}\label{G_0}
	{G}_{{0}}\left(\mathbf{r}\right) \sim r^{-1/2} \left[\sin r - \frac{i\omega}{\omega^{\prime}}\cos r\right],
	\end{equation}
	and
	\begin{equation}
	F_{0}^{s}(\mathbf{r})
	= \tilde{F}^{s}_{0}\left(\mathbf{r}\right)
	\sim \frac{i \Delta}{\omega^{\prime}} r^{-1 / 2} \cos r\label{F0s}, \\
	\end{equation}
	%&F_{0}^{d}(\mathbf{r})=-\left(\tilde{F}_{0}^{d}(\mathbf{r})\right)^{\star} \sim \frac{i \tilde{\Delta}}{\omega^{\prime}} r^{-1 / 2} \cos r e^{2 i \theta}\label{F0d}
	where $ r \equiv |\mathbf{r}| $ and $ \omega^{\prime} \equiv \sqrt{\omega^2 - \Delta^2} $. To second order in $ U_{0} $, we find from Eqs.~(\ref{Green}), (\ref{G_0}), and (\ref{F0s})
	\begin{equation}\label{simulation_s}
	\begin{split}
	&N^{\text{loop},s}(\mathbf{r}) \sim   (r_1r_2d)^{-1/2}\times\\
	&\left\{
	\begin{array}{lr}
	\cos(r_1 + r_2 + d), & \textrm{NM}  \\
	\cos(r_1 + r_2 + d) + \frac{4{\Delta}^{2}}{\omega^{\prime 2}}\cos r_1 \cos r_2 \cos d, & \textrm{M}
	\end{array}
	\right.
	\end{split}
	\end{equation}
	where $ r_{1} = |\mathbf{r} - \mathbf{r}_{1}|$, $ r_{2} = |\mathbf{r} - \mathbf{r}_{2}| $, and $ d $ denotes the distance between the two impurities. 
	The non-magnetic (NM) expression {explains} the elliptical oscillations in Fig.~\ref{s-wave} (b) { (we perform numerical simulations based on the full tight-binding Hamiltonian, details in \cite{numerical_detail})}, where the constant contours are $ r_1 + r_2 = C$, with $C$ given by some constant. In contrast, the magnetic (M) expression is the summation of oscillations of $ r_1 + r_2 $ and that of $ r_1 - r_2 $. As $ r_1 - r_2 = C $ is the very definition of a hyperbola, HF appear in Fig.~\ref{s-wave} (c) on top of the elliptical oscillations. {
	%Intuitively speaking, the emergence of HF is due to the combined effect of magnetic impurities (whose scattering potential takes opposite signs for opposite spins) and the coherence of electrons and holes in the SC (which mixes the two spin species.)
	The emergence of HF is attributed to the loop quasiparticle scattering. { This pattern in the LDOS remains qualitatively unchanged upon including higher-order scattering terms, as described in detail in the Supplementary Material (SM).}
	When $\omega \gg \Delta$, the superconducting coherence is lost on the equal energy contour, and we recover the non-magnetic results (Eq.~(\ref{simulation_s})) which is the same as that in a metal.} {  Also, note that the elliptical and hyperbolic fringes exhibit manifestly different Fourier transformations ((b) (c) inset), which we detailed in the SM.}
	%Adding up higher order scattering processes does not destroy the HF oscillation pattern, since multi-scattering only modifies the factor $ G_{0}(\mathbf{r}_1 - \mathbf{r}_2) $ ($ G_{0}(\mathbf{r}_2 - \mathbf{r}_1) $) in the loop contribution Eq.~(\ref{loop}). As this does not influence the interference depending on $ \mathbf{r} - \mathbf{r}_1 $ and $ \mathbf{r} - \mathbf{r}_2 $, higher order contributions tend to interfere constructively with contributions from lower order.

	%\begin{widetext}
	%\end{widetext}
	
	{\it Chiral SC.---} While some aspects of the HF signal for the $s$-wave minimal model above carry over to chiral pairing symmetries, the twin QPI pattern of a chiral SC already exhibits an HF signal for non-magnetic impurities. Picking $\Delta_{\mathbf{k}}^{p}=\Delta (\sin k_x + i \sin k_y)$ for illustration, we find ${G}_{{0}}$ according to Eq.~(\ref{G_0}) along with
	
		\begin{align}
	&F_{0}^{p}(\mathbf{r})
	=(\tilde{F}_{0}^{p}\left(\mathbf{r}\right))^{\star}
	\sim \frac{i \tilde{\Delta}}{\omega^{\prime}} r^{-1 / 2} \sin r \ e^{i \theta}\label{F0p},
	\end{align}
    where $ \theta $ is the polar angle of $ \mathbf{r} $. This gives a loop contribution to the LDOS according to
	\begin{equation}\label{simulation_p}
	\begin{aligned}
	&N_{p}^{\text {loop }}(\mathbf{r}) \sim\left(r_{1} r_{2} d\right)^{-1 / 2} \times\left\{\cos \left(r_{1}+r_{2}+d\right)+\right. \\
	&\frac{{\Delta}^{2}}{\omega^{\prime 2}}\left[\sin r_{1} \sin r_{2} \cos d \cos \left(\theta_{1}-\theta_{2}\right) + \cos r_{1} \cos r_{2} \cos d\right. \\
	&\left.\pm(\cos r_{1} \sin r_{2} \sin d \cos \theta_{2}-\sin r_{1} \cos r_{2} \sin d \cos \theta_{1})]\right\},
	\end{aligned}
	\end{equation}
	%\begin{equation}\label{simulation_d}
	%\begin{aligned}
	%&N_{d}^{\text {loop }}(\mathbf{r}) \sim\left(r_{1} r_{2} d\right)^{-1 / 2} \times\{\cos \left(r_{1}+r_{2}+d\right)+ \\
	%&\frac{{\Delta}^{2}}{\omega^{\prime 2}}\cos r_1 \cos r_2 \cos d \left[1 + \cos 2(\theta_1 - \theta_2) - \cos 2\theta_1 - \cos 2\theta_2\right]\}
	%\end{aligned}
	%\end{equation}
	where $ \theta_1 $ and $ \theta_2 $ denote the polar angle of $ \mathbf{r} - \mathbf{r}_1 $ and $ \mathbf{r} - \mathbf{r}_2 $, respectively, and $+$/$-$ corresponds to NM/M impurities. According to trigonometric transformations, we immediately recognize that Eq. (\ref{simulation_p}) is also the summation of the oscillations of $ r_1 + r_2 $ and of $ r_1 - r_2 $, albeit the amplitude is further modulated by terms related to the polar angles $ \theta_1 $ and $ \theta_2 $. That is why we also see HF on top of elliptical fringes in Fig.~\ref{loop_p} (b).
	We employ Eq.~(\ref{simulation_p}) in Fig.~\ref{loop_p} (d) to compare our analytical approximation against the numerical calculation in Fig.~\ref{loop_p} (b), which shows good agreement.
	{In contrast to the uniform $s$-wave pairing, here the two scattering processes $\mathbf{r}\rightarrow\mathbf{r}_{1}\rightarrow\mathbf{r}_{2}\rightarrow\mathbf{r}$ and $\mathbf{r}\rightarrow\mathbf{r}_{2}\rightarrow\mathbf{r}_{1}\rightarrow\mathbf{r}$, which are time-reversed to each other, can also interfere to generate the HF, as the superconducting gaps acquire different winding phases on the two paths. This interference is reflected in the $\cos \left(\theta_{1}-\theta_{2}\right)$ term in Eq.~(\ref{simulation_p}).} {We further study the HF signal for a $d+id$-wave SC and find the overall pattern to be similar to that of the $p+ip$-wave case (see SM). The crucial difference is that the interference pattern in the region between two impurity sites for the $d+id$-wave pairing is not uniform but stripe-like, which is attributed to the distinct phase winding of two SC orders and can be explained from their detailed analytic form (details in SM).} {For magnetic impurities, the interference patterns are similar, as detailed in the SM.}
	
	Note that for our analytical expressions Eq.~(\ref{simulation_s}) and~(\ref{simulation_p}) we have assumed $\eta \rightarrow 0^{+}$, while in numerical calculations $\eta$ has to take a non-zero value (which we have chosen as $\eta = 5\times 10^{-3}$). Having a non-zero $\eta$ regulator just amounts to setting $\omega \rightarrow \omega + i \eta$ in Eq.~(\ref{G_0}),~(\ref{F0s}) and~(\ref{F0p}), and does not affect the main features of the LDOS distribution. Departing from the band edge limit where we can compare against our analytical solution, we find that the numerical HF signature prevails even for non-circular Fermi surfaces (Fig.~\ref{loop_p} (c)). This allows to conjecture that the HF pattern is a generic signal for twin-impurity QPI patterns of chiral SCs, where the HF signal dependence on the chemical potential is further detailed in \cite{supp}.

\begin{figure*}[t]
		\centering
		\includegraphics[width=0.7\textwidth]{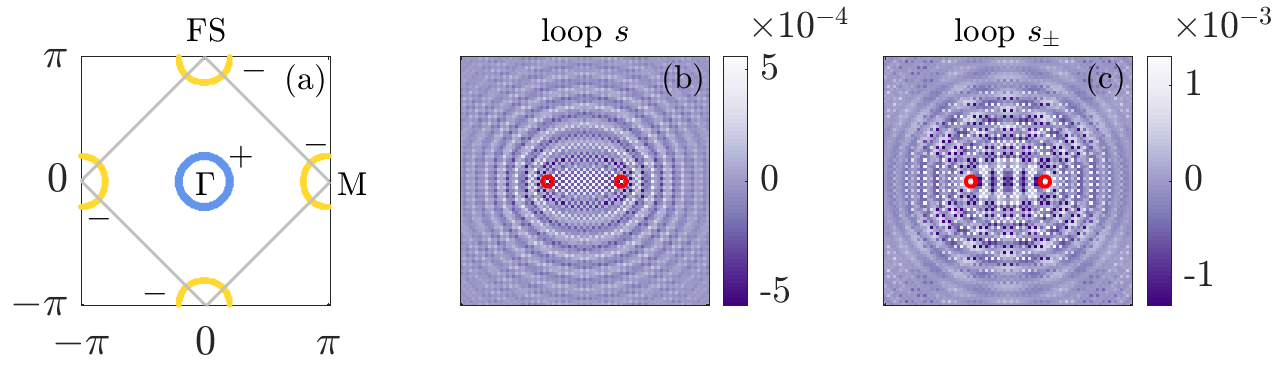}
		\caption{Twin impurity QPI for multi-pocket $s_{\pm}$-wave SC ($ \omega = 0.3232 $, $d = 24$). (a) FS of the ``c" (denoted by yellow line) and the ``d" (denoted by blue line) electron bands of a two-band model. $+$ ($-$) denote the sign of the gap for $s_{\pm}$-wave pairing.
		(b) Loop contribution from two non-magnetic impurities for conventional on-site $ s $-wave pairing around two non-magnetic impurities. (c) Loop contribution as in (b) for $s_{\pm}$-wave pairing.}
		\label{iron}
	\end{figure*}

	\textit{$s_{\pm}$-wave in multi-pocket systems.---}
    Multi-pocket Fermiologies can qualitatively differ from single-pocket systems. In particular, with respect to gapped SCs, multi-pocket systems can exhibit $s_{\pm}$-wave unconventional pairing where the gap function takes opposite signs on different Fermi pockets~\cite{iron_expm}. 
    %The most prominent example is the iron-based high-$T_c$ SC, where it is proposed that the pairing symmetry is the unconventional extended $s$ ($s_{\pm}$)-wave (\cite{iron_expm}). Here we explore whether the rich information contained in the QPI patterns around two impurities is sensitive to sign change in the order parameter, with the example of iron-based SC.
	We start from a two-band model for iron-based SC (\cite{Sykora})
    \begin{equation}
        \begin{aligned}
        H=& \sum_{\mathbf{k}, \sigma}\left(\varepsilon_{\mathbf{k}}^{c} c_{\mathbf{k}, \sigma}^{\dagger} c_{\mathbf{k}, \sigma}^{}+\varepsilon_{\mathbf{k}}^{d} d_{\mathbf{k}, \sigma}^{\dagger} d_{\mathbf{k}, \sigma}\right)^{} \\
        &+\sum_{\mathbf{k}}\left[\Delta_{\mathbf{k}}\left(c_{\mathbf{k}, \uparrow} c_{-\mathbf{k}, \downarrow}+d_{\mathbf{k}, \uparrow} d_{-\mathbf{k}, \downarrow}\right)+\text{h.c.}\right].
        \end{aligned}
    \end{equation}
    Here, ``c" and ``d" label the two electron bands describing an electron-like FS centered at the M point in the folded Brillouin zone (or $(\pi, 0) /(0, \pi)$ for the unfolded zone~\cite{Ding_2008}) and a hole-like FS at the $\Gamma$ point. It reads
    \begin{align}
        &\varepsilon_{\mathbf{k}}^{c}= 2 t\left[\cos \left(k_{x}+k_{y}\right)+\cos \left(k_{x}-k_{y}\right)\right]-\mu_{c}\label{epsilon_c},\\
        &\varepsilon_{\mathbf{k}}^{d}= 2 t\left(\cos k_{x}+\cos k_{y}\right)-\mu_{d}\label{epsilon_d},
    \end{align}
    where we choose $\mu_c = -3.2$ and $\mu_d = 3.6$ (Fig.~\ref{iron}(a)). It turns out that $s_\pm$-wave order, as opposed to conventional $s$-wave order without a sign-changing order parameter, exhibits an HF signal for any kind of impurity. Consider two types of $s$-wave pairing
    \begin{equation}\label{iron_gap}
	\Delta_{\mathbf{k}} = 
	\left\{
	\begin{array}{lr}
	\Delta_{s}, & \  s\textrm{-wave}  \\
	\Delta_{s_{\pm}} \cos k_x \cos k_y, &  \ s_{\pm} \textrm{-wave}
	\end{array}
	\right.
	\end{equation}
	where $t= 1$, $\Delta_{s} = 0.32$, and $\Delta_{s_{\pm}}= 0.4$. The extended $s$-wave pairing gives the same gap magnitude but takes opposite signs on the two FS of the two electron bands.
	%(for more on the gap structure see ref.~\cite{iron_expm, Sykora}). 
	%\begin{widetext}
%\end{widetext}
	The formalism laid out for the single pocket case readily generalizes to the two-band model by recognizing $\mathbf{G}_{{0}}\left(\mathbf{r}\right) = \mathbf{G}^{c}_{{0}}\left(\mathbf{r}\right) + \mathbf{G}^{d}_{{0}}\left(\mathbf{r}\right)$, where $\mathbf{G}^{c}_{{0}}\left(\mathbf{r}\right)$ and $\mathbf{G}^{d}_{{0}}\left(\mathbf{r}\right)$ are defined via Eq.~(\ref{Green}) (the unfolded Brillouin zone is used to perform the summation in the Fourier transformation) and Eqs.~(\ref{epsilon_c})-(\ref{iron_gap}). In Fig.~\ref{iron} (b) and (c), we show the loop contribution in the LDOS distribution of the two-band SC around two non-magnetic impurities for conventional $s$-wave pairing and unconventional $s_{\pm}$-wave pairing, respectively. For Fig.~\ref{iron} (c), a uniform gap ($\tilde{\Delta} = 0.32$) with opposite sign on the two electron bands is used in the underlying calculation in order to reduce the gap anisotropy. Similar to our previous results for conventional $s$-wave pairing on a single pocket, the loop contribution results in elliptical fringes. In addition, however, an extended $s_{\pm}$-wave SC order parameter gives rise to the emergence of HF signal on top of the ellipses. {We note that the emergence of HF is due to the fact that a sign-reversing scattering at a non-magnetic impurity is mathematically equivalent to a sign-preserving scattering at a magnetic impurity. As we have shown for single-pocket $s$-wave SC, the latter case can generate HF.} The ``Mosaic"-like feature (zigzag between neighboring sites) is due to the FS of the electron-like band centering at the M point. Since the FS of the two band model is well approximated by circles, the formulas Eq.~(\ref{loop})-(\ref{F0s}) carry over to the two-band case and analytically explain the HF signal difference between conventional and unconventional $s$-wave pairing \cite{supp}. The latter distinguishability only holds for non-magnetic impurities, as magnetic impurities imply HF pattern in both cases. 
    Our findings for the two-pocket case with sign-changing should generalize to the generic multi-pocket case with sign-changing gap function on individual pockets.

	\textit{Experimental detectability.---}
	The loop contribution is generally smaller than the single contribution according to $ N^{\text{loop}} / N^{\text{single}} \sim (k_F d)^{-1/2} U_{0} / t $ for non-magnetic impurities. {Thus, it is crucial to isolate the loop contribution in STM measurements from the single contribution background, a procedure for which we provide two proposals. The first option involves measuring the LDOS around a single impurity before measuring close to twin impurities. By subtracting the single-impurity contribution from the full contribution, the loop contribution can be isolated. Our second proposal is specifically designed for magnetic impurities, where two measurements are made with the twin impurities with anti-aligned and aligned magnetic moments. While this change does not affect the individual contributions (second order), the loop signal changes sign between the configurations. Taking the difference between the two signals thus significantly suppresses the single contribution and singles out the loop contribution. In the SM, we provide detailed simulations demonstrating the effectiveness of these proposals.}
	%in revealing the hyperbolic fringe patterns from the loop contribution in experiments. }
	%Given twin impurity QPI data from STM, it appears most feasible to subtract the single impurity background to reveal the features of the loop contribution. 
	In the measurements,
	%ideally, a pair of impurities can be found on the surface, which are close to each other but separated from others. In many cases,
	such twin impurities might be fabricated through STM atomic manipulation~\cite{Hla_2014}, which would further allow to explore the evolution of the HF pattern as a function of twin impurity. A simple proof-of-principle experiment for our predicted HF pattern appears to be a conventional $s$-wave superconductor with a clean surface and highly adjustable magnetic impurities on top of it. A good candidate is NbSe$_2$, which can be grown by molecular beam epitaxy yielding clean surfaces~\cite{RN274} required for quasiparticle interference experiments~\cite{PhysRevLett.114.037001}.
	Moreover, with their $s_\pm$-wave pairing, it is also promising to examine the HF signal in iron-based superconductors such as LiFeAs~\cite{doi:10.1126/science.1218726,PhysRevB.84.235121}.

    \textit{Conclusion and outlook.---}
    The hyperbolic fringe (HF) pattern from twin impurity QPI provides a complementary approach of detecting phase information associated with SC order. While conventional pairing necessitates magnetic twin impurities to generate an HF signal, chiral SC and $s_{\pm}$-wave multi-pocket SC already exhibit HF signal for non-magnetic impurities. Upon closer inspection of the principal underlying mathematical structure, it is apparent that the HF signal is not special to SC, but can potentially appear in entirely different contexts such as topological insulators (TIs). For instance, the Qi-Wu-Zhang model on a two-dimensional lattice (\cite{PhysRevB.74.085308}) for time-reversal symmetry breaking TIs exhibits a {\textit{k}-space} Hamiltonian that {has the same form as}
    %is form invariant to
    a chiral $p$-wave SC. It is hence plausible that the HF signature of twin impurities should also appear in the loop contribution to the TI LDOS distribution in the topologically non-trivial regime. For other TIs and topological materials such as graphene, the QPI pattern on the surface around twin impurities presents itself as a valuable perspective for future study.

	\textit{Acknowledgments.---} 
	R.T. thanks his mentor and friend Shoucheng Zhang for a discussion which has provided the inspiration for this project. This work is funded by the Deutsche Forschungsgemeinschaft (DFG, German Research Foundation) through Project-ID 258499086 - SFB 1170 and through the W\"urzburg-Dresden Cluster of Excellence on Complexity and Topology in Quantum Matter - ct.qmat Project-ID 390858490 - EXC 2147. This work is also supported by the Singapore National Research Foundation QEP grant (Grant No. NRF2021-QEP2-02-P09).

	%	\nocite{*}
	%	\bibliographystyle{unsrt}
	\bibliography{quasi_ref}

\end{document}

\title{\textbf{Supplemental material for ``Hyperbolic fringe signal for twin impurity quasiparticle interference"}}
    
	\author{Peize Ding}
	\email{pd2714@columbia.edu}
	\affiliation{Institute for Theoretical Physics,
		University of W\"{u}rzburg, Am Hubland, D-97074 W\"{u}rzburg, Germany}
	\affiliation{School of the Gifted Young, University of Science and Technology of China, Hefei 230026, China}
	\affiliation{Department of Physics, Columbia University, New York 10027, NY, USA}
	\author{Tilman Schwemmer}
	\affiliation{Institute for Theoretical Physics,	University of W\"{u}rzburg, Am Hubland, D-97074 W\"{u}rzburg, Germany}
	\author{Ching Hua Lee}
	\affiliation{Department of Physics, National University of Singapore, Singapore, 117542}
	\affiliation{Joint School of National University of Singapore and Tianjin University, International Campus of Tianjin University, Binhai New City, Fuzhou 350207, China}
	\author{Xianxin Wu}
	\email{xxwu@itp.ac.cn}
	\affiliation{CAS Key Laboratory of Theoretical Physics, Institute of Theoretical Physics, Chinese Academy of Sciences, Beijing 100190, China}
    %\affiliation{Max-Planck-Institut f\"ur Festk\"orperforschung, Heisenbergstrasse 1, D-70569 Stuttgart, Germany}

	\author{Ronny Thomale}
	\email{rthomale@physik.uni-wuerzburg.de}
	\affiliation{Institute for Theoretical Physics,
		University of W\"{u}rzburg, Am Hubland, D-97074 W\"{u}rzburg, Germany}
	
	\maketitle
	
	\onecolumngrid
	
	The Supplementary material contains the following content arranged by sections:
	
	\begin{itemize}
	    \item Derivation of Eq.~(9)-(11) in the main text (the Green's function in real space) by performing Fourier transformation.
	    \item Discussion of twin impurity QPI for $d+id$-wave SC, in parallel with that of $p+ip$-wave SC in the main text.
	    \item Discussion about the robustness of HF away from band bottom.
	    \item Analysis of the patterns of HF in $p+ip$-wave and $d+id$-wave SC.
	    \item Discussion of the mathematical origin of HF in iron-based SC.
	    \item Analytical investigation of higher order loop contribution.
	    \item Exhibition of patterns of HF around magnetic impurities.
	    \item Experimental proposal to detect HF in iron-based SC.
	    \item A detailed analysis of the Fourier spectra of the twin impurity QPI signals.
	\end{itemize}
	
    \section*{I. Green's function integrals}
    
	In this section, we evaluate the Green's function assuming circular Fermi surface by performing Fourier transformation. The normal Green's function in real space reads
	
	\begin{equation}\label{key}
	\begin{split}
	G_{0}\left(\mathbf{r}, \omega\right) &= 
	\int \frac{d\mathbf{k}}{(2\pi) ^ 2}
	e^{i \mathbf{k}\mathbf{r}}
	\frac{\omega + \varepsilon_{\mathbf{k}}}{\omega ^ 2 - \varepsilon ^ 2_{\mathbf{k}} - |\Delta_{\mathbf{k}}|^2 + i 0^{+}}
	%	 = \frac{1}{(2\pi)^2}\int_{0}^{\infty}k dk
	%	\int_{0}^{2\pi}d\phi e^{ikr\cos\phi}
	%	\frac{\omega + \varepsilon_{\mathbf{k}}}{\omega ^ 2 - \varepsilon ^ 2_{\mathbf{k}} - |\Delta_{\mathbf{k}}|^2 + 2\omega i \eta}
	= -\frac{1}{8\pi^2ta^2} \int_{-\mu^{\prime}}^{\infty} d\xi
	\int_{0}^{2\pi}d\phi e^{ir\cos\phi \sqrt{2m(\xi + \mu^{\prime})}}
	\frac{\omega + \xi}{\xi^2 -\omega^2 + \Delta^ 2 - i 0^{+}}.
	\end{split}
	\end{equation}
	Here $ \varepsilon_{\mathbf{k}} = \frac{k^2}{2m} - \mu^{\prime} $ ($\hbar = 1$) where $ \mu^{\prime} = \mu + 4t $ is the chemical potential measured from the band bottom, and $\phi$ is the polar angle of $\mathbf{k}$ measured relative to $\mathbf{r}$. We have used the continuous version of $ \varepsilon_{\mathbf{k}} $ near the band bottom, and set the range of $ \mathbf{k} $ integration to be $ R^2 $. We perform the integration assuming $ \omega > \Delta $, and $ \mu^{\prime} \gg \sqrt{\omega^2 - \Delta^2} $, which is the case in our numerical study.
	Note that the integrand decays quickly so that we can set $ -\mu^{\prime} $ to $ -\infty $ in the lower limit. We evaluate
	\begin{equation}\label{key}
	\begin{split}
	I_{0} = 
	\int_{-\infty}^{\infty} d\xi
	\int_{0}^{2\pi}d\phi e^{ir\cos\phi \sqrt{2m(\xi + \mu^{\prime})}}
	\frac{1}{\xi^2 - \omega^2 + \Delta^ 2 - i 0^{+}}
	\approx \frac{2\pi^2 i}{\sqrt{\omega^2 - \Delta ^ 2}} J_{0}(k_F r)
	\end{split}
	\end{equation}
	\begin{equation}\label{key}
	\begin{split},
	I_{1} = 
	\int_{-\infty}^{\infty} d\xi
	\int_{0}^{2\pi}d\phi e^{ir\cos\phi \sqrt{2m(\xi + \mu^{\prime})}}
	\frac{\xi}{\xi^2 -\omega^2 + \Delta^ 2 - i 0^{+}} 
	\frac{\omega_D^2}{\xi^2 + \omega_D^2}
	= -{2\pi^2} H_{0}(k_F r).
	\end{split}
	\end{equation}
	We have introduced the convergence factor due to the Debye cutoff, and we have taken $ \omega_D $ to infinity (because $ \Delta \ll \omega_D $) at last \cite{helical_shiba}. Also, we have made use of the relation $ \mu^{\prime} \gg \sqrt{\omega^2 - \Delta^2} $, and,
	\begin{equation}\label{key}
	\int_{0}^{\pi}d\phi e^{ix\sin\phi} = \pi\left[J_{0}(x) + i H_{0}(x)\right],
	\end{equation}
	where $ J_{0}(x), H_{0}(x) $ denotes Bessel and Struve function, respectively.
	Finally, using the asymptotic form of Bessel and Struve function when $k_F r\gg 1$, we have
	\begin{equation}\label{G_sp}
	\begin{split}
	G_{0}(\mathbf{r}) = \frac{1}{4} \left[ H_{0}(k_F r) - i\frac{\omega}{\sqrt{\omega^2 - \Delta ^ 2}} J_{0}(k_F r)\right]
	\approx (8\pi k_F r)^{-\frac{1}{2}} \left[\sqrt{\frac{2}{\pi k_F r}} + \sin(k_F r - \frac{\pi}{4}) - i\frac{\omega}{\sqrt{\omega^2 - \Delta ^ 2}}\cos(k_F r - \frac{\pi}{4})\right].
	\end{split}
	\end{equation}
	For $ s $-wave pairing, we simply have
	\begin{equation}\label{F_s}
	\begin{split}
	F_{0}^{s}\left(\mathbf{r}\right) = \tilde{F}^{s}_{0}\left(\mathbf{r}\right) =
	\int \frac{d\mathbf{k}}{(2\pi) ^ 2}
	e^{i \mathbf{k}\mathbf{r}}
	\frac{\Delta}{\omega ^ 2 - \varepsilon ^ 2_{\mathbf{k}} - |\Delta_{\mathbf{k}}|^2 + i 0 ^{+}}
	= -\frac{i\Delta}{4\sqrt{\omega^2 - \tilde{\Delta }^ 2}}J_{0}(k_F r)
	= -(8\pi k_F r)^{-\frac{1}{2}} \frac{i \Delta}{\sqrt{\omega^2 - \Delta ^ 2}} \cos(k_F r - \frac{\pi}{4})
	\end{split}
	\end{equation}.
	While for $ p+ip $-wave pairing, we have
	\begin{equation}\label{F_p}
	\begin{split}
	F_{0}^{p}\left(\mathbf{r}\right)  =(\tilde{F}_{0}^{p}\left(\mathbf{r}\right))^{\star} 
	&=
	\int \frac{d\mathbf{k}}{(2\pi) ^ 2}
	e^{i \mathbf{k}\mathbf{r}}
	\frac{\tilde{\Delta} e^{i(\theta + \phi)}}{\omega ^ 2 - \varepsilon ^ 2_{\mathbf{k}} - |\Delta_{\mathbf{k}}|^2 + i 0^{+}}\\
	&= \frac{i\tilde{\Delta}}{4\sqrt{\omega^2 - \tilde{\Delta }^ 2}}J_{1}(k_F r)e^{i\theta}
	\approx\frac{\tilde{\Delta}}{\sqrt{\omega^2 - \tilde{\Delta} ^ 2}} (8\pi k_F r)^{-\frac{1}{2}} \sin(k_F r - \frac{\pi}{4}) i e^{i\theta},
	\end{split}
	\end{equation}
	where $ \tilde{\Delta} $ is the gap magnitude (assuming uniform) on the Fermi surface. The first equality can be read off from symmetry argument, and $\theta$ denotes the polar angle of $\mathbf{r}$.
	Similarly, for chiral $ d $-wave pairing
	\begin{equation}\label{F_d}
	\begin{split}
	F_{0}^{d}(\mathbf{r})= -(\tilde{F}_{0}^{d}\left(\mathbf{r}\right))^{\star} 
	&=
	\int \frac{d\mathbf{k}}{(2\pi) ^ 2}
	e^{i \mathbf{k}\mathbf{r}}
	\frac{\tilde{\Delta} e^{2i(\theta + \phi)}}{\omega ^ 2 - \varepsilon ^ 2_{\mathbf{k}} - |\Delta_{\mathbf{k}}|^2 + i 0^{+}}\\
	&= -\frac{i\tilde{\Delta}}{4\sqrt{\omega^2 - \tilde{\Delta }^ 2}}J_{2}(k_F r)e^{2i\theta}
	\approx \frac{\tilde{\Delta}}{\sqrt{\omega^2 - \tilde{\Delta }^ 2}} (8\pi k_F r)^{-\frac{1}{2}} \cos(k_F r - \frac{\pi}{4}) i e^{2i\theta}.
	\end{split}
	\end{equation}
	
	\section*{II. Twin impurity QPI for $d+id$-wave SC}
	
	In this section, we discuss twin impurity QPI for $d+id$-wave SC, in parallel with our discussion of $p+ip$-wave SC in the main text.
	
\begin{figure*}[tbp]
		\centering
		\includegraphics[width=0.5\textwidth]{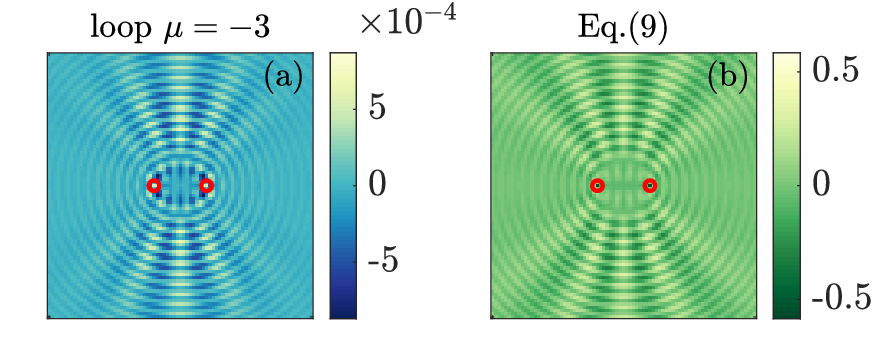}
		\caption{Twin impurity QPI for $d+id$-wave SC. (a) Loop contribution around two non-magnetic impurities. (b) Simulation of Eq.~(\ref{simulation_d}) on a $81\times 81$ real space lattice. $\mu = -3$, $\Delta = 0.05$, $\omega = 1.1 \Delta$ and $d = 16$ are used. HF appear in $d+id$-wave SC around non-magnetic impurities, in consistence with our discussion about $p+ip$-wave SC. Our analytical formula (b) agrees with numerical result (a).}
		\label{d-wave}
	\end{figure*}

	The gap function we adopt is $\Delta_{\mathbf{k}} = {2\Delta}(\cos k_x - \cos k_y + i \sin k_x \sin k_y) $. In Fig.~\ref{d-wave} (a), we show the loop contribution to the LDOS distribution around two non-magnetic impurities for $d+id$-wave pairing (Magnetic impurity case is shown in Sec.~\ref{magneticsec}). Similar to $p+ip$-wave pairing, HF show up regardless of the nature of impurities. In analogy to chiral $p$-wave, we can also derive analytical formulas to understand the pattern when Fermi surface is nearly a circle. Using Eq.~(\ref{G_sp}), (\ref{F_d}) and Eq.~(7) in the main text, we have
	\begin{equation}\label{simulation_d}
	%\begin{aligned}
	N_{d}^{\text {loop }}(\mathbf{r}) \sim\left(r_{1} r_{2} d\right)^{-1 / 2} \times\{\cos \left(r_{1}+r_{2}+d\right)+ 
	\frac{{\Delta}^{2}}{\omega^{\prime 2}}\cos r_1 \cos r_2 \cos d \left[1 + \cos 2(\theta_1 - \theta_2) \mp (\cos 2\theta_1 + \cos 2\theta_2\right)]\},
	%\end{aligned}
	\end{equation}
	where $-$ and $+$ corresponds to NM and M case, respectively.
	In Fig.~\ref{d-wave} (b), we present result from Eq.~(\ref{simulation_d}), which shows a good agreement with numerical data (Fig.~\ref{d-wave} (a)).

	\section*{III. Robustness of HF away from band bottom}
	
	\begin{figure*}[tbp]
		\centering
		\includegraphics[width=1.0\textwidth]{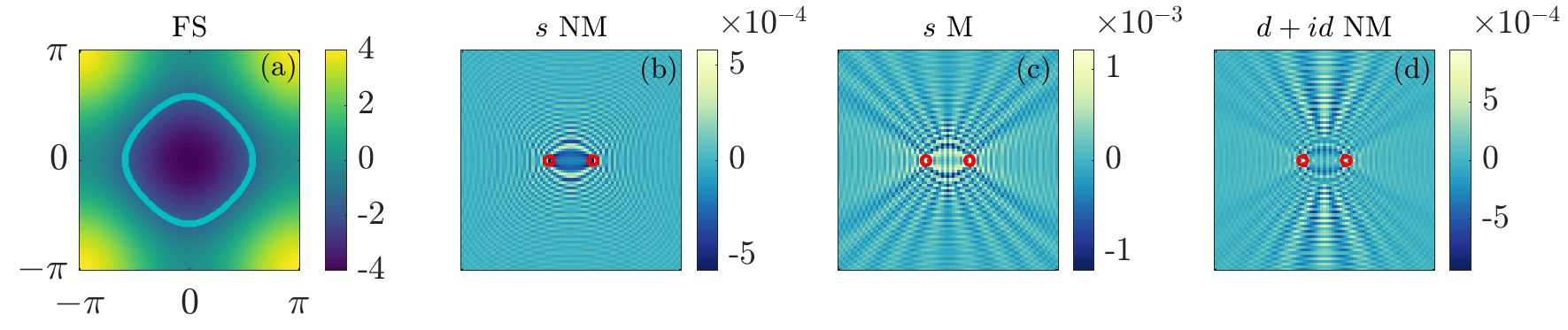}
		\caption{Robustness of HF away from band bottom. (a) Band dispersion $\epsilon(\mathbf{k}) = -2 (\cos k_x + \cos k_y)$ (denoted by color) and the FS at  $\mu = -1.5$ (blue line) in the first Brillouin zone. (b) ((c)) Loop contribution around two non-magnetic (magnetic) impurities for an $s$-wave SC. (d) Loop contribution around two non-magnetic impurities for chiral $d$-wave SC. $\Delta = 0.05$ for $s$-wave and $\Delta = 0.02$ for $d$-wave (to make the average gap magnitude on the Fermi surface similar with $s$-wave). $\mu = -1.5$, $\omega = 1.1 \Delta$ and $d = 16$ are used. HF appear robustly in chiral SC around non-magnetic impurities and in conventional SC around magnetic impurities.}
		\label{robust}
	\end{figure*}	
	
	When the chemical potential is tuned away from band bottom, our analytical formulas no longer hold. However, from numerical calculation we show that the HF is actually robust in chiral $d$-wave SC (the similar plot for chiral $p$-wave SC has been shown in Fig.~2 (c) in the main text) around non-magnetic impurities and in conventional SC around magnetic impurities, as shown in Fig.~\ref{robust} (d) and (c), respectively. For conventional SC around twin non-magnetic impurities (Fig.~\ref{robust} (b)), the HF do not appear, and the elliptical fringes (that appear when FS is circular) are distorted. {Note that the four arrays of curves in Fig.~\ref{robust} (b) that resemble hyperbolas appear as a consequence of distorted Fermi surface. These trivial hyperbolas are distinct from the HF emerging due to non-trivial interference, whose foci are located at the sites of two impurities. }
 %to be not confused with the HF emerging due to non-trivial interference.}
	
	We then demonstrate that the chemical potential value $+\mu$ and $-\mu$ give equivalent results. When $\mu$ is positive, the Fermi surface centers around the $M$ ($(\pi, \pi) \equiv \mathbf{Q}$) point. Hence compared to Green's function with negative $\mu$, there is an additional factor of $e^{i\mathbf{Q}\mathbf{r}}$ in the Green's function with positive $\mu$. However, in the loop contribution, this additional factor cancels out in the product of the three Green's function (Eq.~(8) in the main text). Therefore, with $+\mu$ and $-\mu$ we essentially have equivalent results. Therefore, our results hold for all the chemical potential value within the band width, as long as the Fermi surface is not in the vicinity of van Hove singularities.
	
	%\section{Derivation}
	
%	\begin{equation}
%	\begin{split}
 %   	&G_{0}\left(\mathbf{r}-\mathbf{r}_{1}\right) G_{0}\left(\mathbf{r}_{1}-\mathbf{r}_{2}\right) G_{0}\left(\mathbf{r}_{2}-\mathbf{r}\right)
  %  	\mp F_{0}\left(\mathbf{r}-\mathbf{r}_{1}\right) \tilde{F}_{0}\left(\mathbf{r}_{1}-\mathbf{r}_{2}\right) G_{0}\left(\mathbf{r}_{2}-\mathbf{r}\right)\\
   % 	&\mp G_{0}\left(\mathbf{r}-\mathbf{r}_{1}\right) F_{0}\left(\mathbf{r}_{1}-\mathbf{r}_{2}\right) \tilde{F}_{0}\left(\mathbf{r}_{2}-\mathbf{r}\right)
    %	-F_{0}\left(\mathbf{r}-\mathbf{r}_{1}\right) G_{0}\left(\mathbf{r}_{1}-\mathbf{r}_{2}, -\omega\right) \tilde{F}_{0}\left(\mathbf{r}_{2}-\mathbf{r}\right)
	%\end{split}
    %\end{equation}
    
    \section*{IV. Patterns of HF in $p+ip$-wave and $d+id$-wave SC around NM impurities}\label{analysis_pd}
    
    Both $p+ip$-wave and $d+id$-wave SC can host HF around twin non-magnetic impurities. While the HF look similar, they exhibit distinct patterns due to different phase winding of the order parameter ($\Delta e^{i\theta}$ for $p+ip$-wave and $\Delta e^{2i\theta}$ for $d+id$-wave), which we investigate in detail in this section.
    
    \begin{figure*}[tbp]
		\centering
		\includegraphics[width=1.0\textwidth]{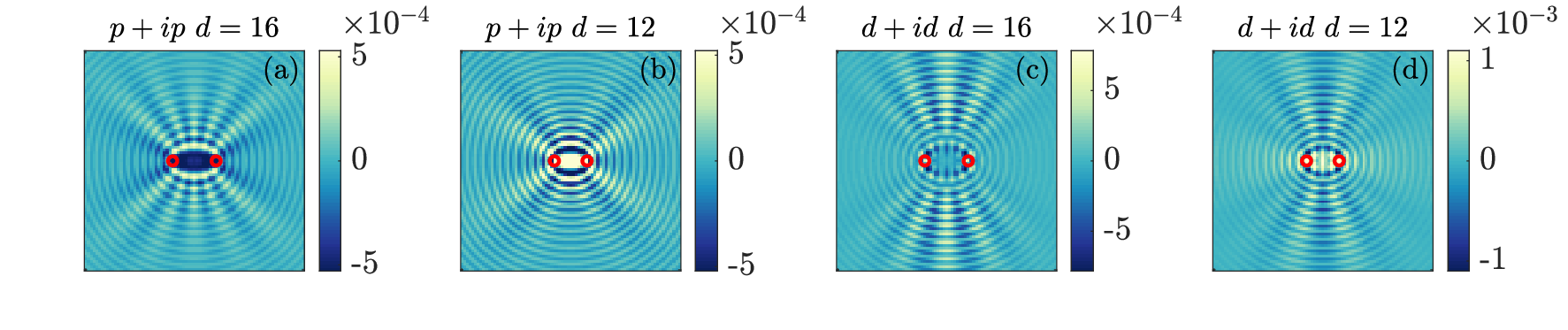}
		\caption{Patterns of HF in $p+ip$-wave and $d+id$-wave SC. (a)((b)) Loop contribution around two non-magnetic impurities in $p+ip$-wave SC for $d = 16$ ($d = 12$). (c)((d)) Loop contribution around two non-magnetic impurities in $d+id$-wave SC for $d = 16$ ($d = 12$). $\mu = -3$, $\Delta = 0.05$ and $\omega = 1.1 \Delta$ are used. There is a constant region in the center in the HF in $p+ip$-wave SC. There is a clear $\pi$-phase shift between the hyperbolic oscillations in the central region and in the outside region in $d+id$-wave SC.}
		\label{compare}
	\end{figure*}
	
	In Fig.~\ref{compare}, we review the loop contribution in $p+ip$-wave and $d+id$-wave SC for $d = 16$, and further show the HF for a different distance $d = 12$. We see that there is a nearly constant region in the center (near the line between the two impurities) in the HF in $p+ip$-wave SC, regardless of the distance. In contrast, the HF in the $d+id$-wave SC exhibit oscillations and is stripe-like in the central region. To understand this crucial difference, we review the analytical formula for loop contribution in $p+ip$-wave SC
	\begin{equation}\label{simulation_p}
	\begin{aligned}
	&N_{p}^{\text {loop }}(\mathbf{r}) \sim\left(r_{1} r_{2} d\right)^{-1 / 2} \times\left\{\cos \left(r_{1}+r_{2}+d\right)+\right. \\
	&\frac{{\Delta}^{2}}{\omega^{\prime 2}}\left[\sin r_{1} \sin r_{2} \cos d \cos \left(\theta_{1}-\theta_{2}\right) + \cos r_{1} \cos r_{2} \cos d\right. \left.
	\pm
	(\cos r_{1} \sin r_{2} \sin d \cos \theta_{2}-\sin r_{1} \cos r_{2} \sin d \cos \theta_{1}]\right)\},
	\end{aligned}
	\end{equation}
	($+/-$ corresponds to the case of NM/M impurities)
	and Eq.~(\ref{simulation_d}) for $d+id$-wave SC. To see what happens in the central region, we take $\theta_1 = \delta$, $\theta_2 = \pi - \delta$, where $\delta$ denotes a small angle. Note also that $r_1 + r_2 \approx d$ in that region. We see that the second term (which is relevant to HF) in Eq.~(\ref{simulation_p}) limits to $\cos{2d}$ at small $\delta$, which is indeed a constant in space. In contrast, in the second term in Eq.~(\ref{simulation_d}), the leading order reads $\cos r_1 \cos r_2 \cos d\ \delta^2$, giving rise to the oscillations (depending on $r_1 - r_2$) in the central region.
	
	In Fig.~\ref{compare} (c) and (d), one can also see a clear $\pi$-phase shift between the hyperbolic oscillations in the central region and in the outside region. To understand this, we focus on the vertical bisector of the two impurities, where $\theta_1 = \pi - \theta_2 = \theta$. The second term in Eq.~(\ref{simulation_d}) is proportional to $\cos r_1 \cos r_2 \cos d\ \varphi(\theta)$, where the modulation factor $\varphi(\theta) = 1 + \cos(4\theta) - 2 \cos(2\theta)$. It is easy to see that $\varphi(\theta) < 0$ when $0<\theta<\pi/4$, and $\varphi(\theta) > 0$ when $\pi/4<\theta<\pi/2$. This sign difference explains the $\pi$-phase shift in the hyperbolic oscillations in the central region.
	
	At last, we briefly mention that, to distinguish between a $p_{x}$-wave and a $d_{x^2-y^2}$-wave \textit{gapless} SC, QPI around single impurity is enough to tell the difference, through different symmetries in the LDOS patterns (\cite{extrema}).

	\section*{V. Mathematical origin of HF in iron-based SC around NM impurities}
	
	As mentioned in the main text, for the two-band model, the Green's function $\mathbf{G}_{{0}}\left(\mathbf{r}\right) = \mathbf{G}^{c}_{{0}}\left(\mathbf{r}\right) + \mathbf{G}^{d}_{{0}}\left(\mathbf{r}\right)$, where $\mathbf{G}^{c}_{{0}}\left(\mathbf{r}\right)$ and $\mathbf{G}^{d}_{{0}}\left(\mathbf{r}\right)$ are defined via Eq.(4) (main text) for the ``c" and ``d" band, respectively. To the lowest order of impurity potential, the loop contribution around non-magnetic impurities reads
	\begin{equation}
	   \mathbf{G}_0(\mathbf{r}-\mathbf{r}_2)U_{\textrm{NM}}\mathbf{G}_0(\mathbf{r}_2-\mathbf{r}_1)U_{\textrm{NM}}\mathbf{G}_0(\mathbf{r}_1-\mathbf{r})+(1\leftrightarrow 2),
	\end{equation}
	where $U_{\textrm{NM}} = U_{0}\tau_{z}$. We claim that it is the $\mathbf{G}^{d}_0(\mathbf{r}-\mathbf{r}_2)U_{\textrm{NM}}\mathbf{G}^{c}_0(\mathbf{r}_2-\mathbf{r}_1)U_{\textrm{NM}}\mathbf{G}^{d}_0(\mathbf{r}_1-\mathbf{r})$ (and also $\mathbf{G}^{c}_0(\mathbf{r}-\mathbf{r}_2)U_{\textrm{NM}}\mathbf{G}^{d}_0(\mathbf{r}_2-\mathbf{r}_1)U_{\textrm{NM}}\mathbf{G}^{c}_0(\mathbf{r}_1-\mathbf{r})$) term that gives rise to HF. To see this, note that, apart from an unimportant phase factor (due to the fact that the ``c" Fermi pocket centers at M-point), the essential difference between $\mathbf{G}^{c}_{{0}}\left(\mathbf{r}\right)$ and $\mathbf{G}^{d}_{{0}}\left(\mathbf{r}\right)$ is that, $F^{c}_{{0}}\left(\mathbf{r}\right) = - F^{d}_{{0}}\left(\mathbf{r}\right)$. Then, it is not hard to see that
	\begin{equation}
	    \mathbf{G}^{d}_0(\mathbf{r}-\mathbf{r}_2)U_{\textrm{NM}}\mathbf{G}^{c}_0(\mathbf{r}_2-\mathbf{r}_1)U_{\textrm{NM}}\mathbf{G}^{d}_0(\mathbf{r}_1-\mathbf{r}) \sim
	    \mathbf{G}^{d}_0(\mathbf{r}-\mathbf{r}_2)U_{\textrm{M}}\mathbf{G}^{d}_0(\mathbf{r}_2-\mathbf{r}_1)U_{\textrm{M}}\mathbf{G}^{d}_0(\mathbf{r}_1-\mathbf{r}), \label{iron_math}
	\end{equation}
	up to a phase factor, where $U_{\textrm{M}} = U_{0}\sigma_{z}$ denotes \textit{magnetic} impurity potential. We immediately recognize that the right-hand side simply amounts to scattering around magnetic impurities in an $s$-wave SC, hence giving rise to HF. If the pairing symmetry in the iron-based SC model is conventional $s$-wave, on the other hand, there is no property as in Eq.~(\ref{iron_math}) hence no HF.
	
	\section{VI. Higher order loop contributions}
	
	In the main text, we investigated the loop contribution to the second order of impurity potential and showed that while $N^{loop, s}(\mathbf{r})$ features elliptical oscillations around NM impurities (Eq.(11) in the main text), HF appear around M impurities for an $s$-wave SC, around NM/M impurities for chiral SC (Eq.(13) in the main text). In reality, multi-scattering processes also contribute to the LDOS signal measured in experiments. This requires us to answer the question from analytical perspectives that, do higher order scattering processes produce an $N^{loop, s}(\mathbf{r})$ that only features elliptical oscillations without HF around NM impurities? Also, do HF also appear in higher order scattering processes around NM impurities in chiral SCs?
	
	We study $N^{loop, s}(\mathbf{r})$ around NM impurities firstly. There are two kinds of multi-scattering processes. One is scattering at the same impurity site multiple times. The other is scattering between two impurities multiple times. No matter which kind, we note that one can write the contribution as $\mathbf{G}_0(\mathbf{r}-\mathbf{r}_2)U_{\textrm{NM}}\Bar{\mathbf{G}}_0(\mathbf{r}_2-\mathbf{r}_1)U_{\textrm{NM}}\mathbf{G}_0(\mathbf{r}_1-\mathbf{r})+(1\leftrightarrow 2)$, where $\Bar{\mathbf{G}}_0(\mathbf{r}_2-\mathbf{r}_1)$ represents the renormalized Green's function incorporating multi-scattering processes, i.e. $\Bar{\mathbf{G}}_0(\mathbf{r}_2-\mathbf{r}_1) = \mathbf{G}_{0}(\mathbf{0})U_{\textrm{NM}}...U_{\textrm{NM}}\mathbf{G}_0(\mathbf{r}_{2}-\mathbf{r}_1) U_{\textrm{NM}}...U_{\textrm{NM}} \mathbf{G}_{0}(\mathbf{0})$. To see what form $\Bar{G}_{0}$ takes, we firstly study $G_{0}(\mathbf{d}_{1}) U_{\textrm{NM}} G_{0}(\mathbf{d}_{2})$, where $\mathbf{d}_{1} = d_{1} \hat{x}$, $\mathbf{d}_{2} = d_{2} \hat{x}$, and $d_{1}, d_{2}$ are two arbitrary real constant. With some algebra, one can show that
	\begin{equation}
	    \mathbf{G}_{0}^{s}(\mathbf{d}_{1}) U_{\textrm{NM}} \mathbf{G}_{0}^{s}(\mathbf{d}_{2})
	    =
	    \sin d^{\prime}\tau_{z} - \frac{i\omega}{\omega^{\prime}}\cos d^{\prime} I + \frac{i \Delta}{\omega^{\prime}} \cos d^{\prime} \tau_{x}
	    = 
	    \mathbf{G}_{0}^{s}(\mathbf{d}^{\prime}),
	\end{equation}
	where $\mathbf{d}^{\prime} = d^{\prime}\hat{x}$ with $d^{\prime} = d_1 + d_2 - \frac{\pi}{2}$. According to this elegant result, $\Bar{G}_{0}$ can be expressed as $G^{s}_{0}(\Bar{d})$ with $\Bar{d}$ a constant, regardless of the complexity of the multi-scattering process. This immediately implies that Eq.~(11) (in the main text) still holds for higher order scattering terms for NM impurities, albeit the parameter $d$ is modulated. Therefore, we successfully explained the absence of HF in higher order contributions around NM impurities for $s$-wave states.
	
	We then move on to the discussion of chiral SC states. After a similar calculation, we can obtain
	\begin{equation}
	\begin{split}
	    \mathbf{G}_{0}^{p+ip}(\mathbf{d}_{1}) U_{\textrm{NM}} \mathbf{G}_{0}^{p+ip}(\mathbf{d}_{2})
	    &=
	    [(1 - \frac{\Delta^{2}}{\omega^{\prime 2}}) \sin d_1 \sin d_2 - \frac{\omega^2}{\omega^{\prime 2}} \cos d_1 \cos d_2]\tau_{z}\\
	    &- \frac{i\omega}{\omega^{\prime}}\sin(d_1 + d_2) I
	    - \frac{\omega \Delta}{\omega^{\prime 2}} \sin(d_1 - d_2) \tau_{x}
	    - \frac{2\Delta}{\omega^{\prime}} \sin d_1 \sin d_2 \tau_{y}.
	\end{split}
	\end{equation}
	This time it cannot be rewritten into a normalized Green's function as the case of $s$-wave. Nevertheless, one can easily show that, such a complex $\Bar{G}_{0}(\mathbf{r}_2 - \mathbf{r}_1)$ also gives rise to HF as in Eq.~(13) (in the main text), where the spatial frequency of oscillation is still given by $1 / k_{F}$. Therefore, our lowest order calculation in the main text can capture the main feature of the total loop contribution. Similar arguments hold for scattering around M impurities in an $s$-wave SC and scattering around NM/M impurities in chiral $d$-wave SC.
	
\begin{figure*}[tbp]
		\centering
		\includegraphics[width=0.5\textwidth]{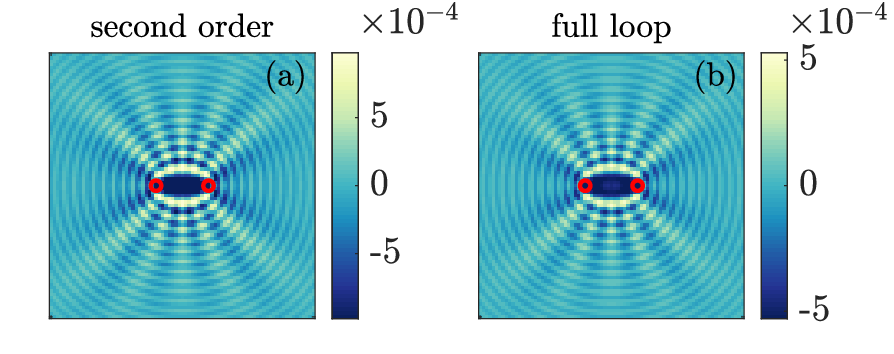}
		\caption{Numerical comparison of the second order loop contribution (a) and the full loop contribution (b), around non-magnetic impurities in a chiral p-wave state. $d=16, \Delta=0.05, \omega=1.1\Delta$ are used.}
		\label{compare_second}
	\end{figure*}

	At last, we emphasize that the figures shown in the main text are calculated numerically where {\it all the higher order contributions are summed up}. To make our arguments even more compelling, we provide a numerical comparison of the second order loop contribution and the full loop contribution, around non-magnetic impurities in a chiral p-wave state, in Fig.~\ref{compare_second}. Despite some quantitative difference in the intensity, they apparently exhibit the same qualitative pattern, ensuring us that the second order approximation captures the key features.
	
	\section{VII. Patterns of HF around magnetic impurities}\label{magneticsec}
	
\begin{figure*}[tbp]
		\centering
		\includegraphics[width=1.0\textwidth]{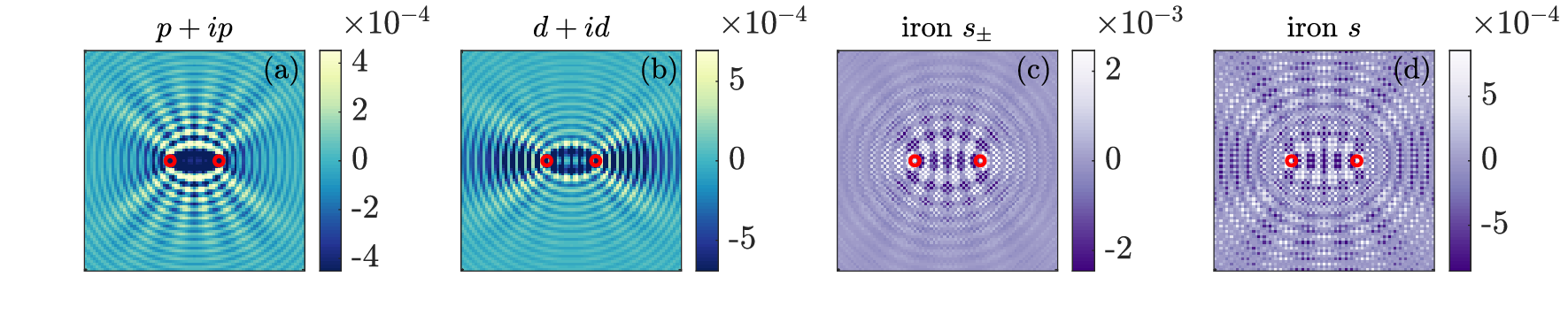}
		\caption{Patterns of HF around magnetic impurities in $p+ip$-wave (a) and $d+id$-wave SC, and in $s_{\pm}$-wave (c) and $s$-wave (d) state for the iron based model (see Eq.~(14) and Eq.~(17) in the main text). For (a) and (b), $\Delta = 0.05, \omega = 1.1 \Delta, \mu = -3, d = 18$. For (c) and (d), the same parameters are used as in Fig.~3 in the main text. Note that the distinct characteristic can also be used to tell these pairing symmetries apart experimentally.}
		\label{magnetic}
	\end{figure*}

	To facilitate a comprehensive analysis of our work, we have depicted the patterns HF in the vicinity of magnetic (M) impurities for various SC states, in Fig.~\ref{magnetic}. Of particular note is the contrasting behavior exhibited by the HF in $p+ip$ and $d+id$-wave pairing states. The HF signals in the $p+ip$-wave state exhibit a constant plateau in the region that lies between the two impurities, while those in the $d+id$-wave state are nearly absent along the vertical bisector of the impurities. Note that these features can be explained on a theoretical footing through our analytical formulas Eq.~\ref{simulation_p} and \ref{simulation_d}, in the same spirit as in Sec.~\ref{analysis_pd} via taking limits.

    Furthermore, the HF patterns around M impurities are distinctive in their characteristics. Despite the presence of HF in both $s_{\pm}$-wave and normal $s$-wave pairing in iron-based superconductors, the HF signals exhibit significant diversity (as evident from the comparison between Fig.~\ref{magnetic} (c) and (d)). As such, the HF patterns around M impurities can serve as an indicator for identifying the presence of sign-changing order parameters.
	
	\section*{VIII. Experimental detection of HF in iron-based SC}

	In this section, we estimate the magnitude of the loop contribution using realistic parameters of iron-based SC. In the main text, we mentioned that the loop contribution is smaller than the single contribution by $ N^{\text{loop}} / N^{\text{single}} \sim (k_F d)^{-1/2} U_{0} / t $ for non-magnetic impurities. For a typical iron-based SC, $a \approx 4$~\AA, $k_F \approx 0.2 \pi / a$ (\cite{Ding_2008,iron_expm}), hopping $t \approx 0.3 $ eV, and a typical scattering potential $U \approx 0.2$ eV (\cite{PhysRevB.82.224506}). For a pair of twin impurities with a distance of $d= 24a$ as in the main text, $ N^{\text{loop}} / N^{\text{single}} \approx 1 / 6 $. The magnitude of the single contribution is about $0.01 (ta^2)^{-1}$, hence the magnitude of the loop contribution is roughly $0.001 (ta^2)^{-1}$. On the other hand, a typical STM experiment nowadays has a high energy resolution of 20 $\mu$eV and a spatial resolution of 0.1 nm ~\cite{naturematerial}. Converting to our units (where lattice spacing $a$ and hopping $t$ is set to unity), typical LDOS resolution (in unit of $(ta^2)^{-1}$) is at least $10^{-5}$. 
	%Given the band structure and gap amplitude, it is easy to calculate the single contribution to high accuracy. Subtracting that from the detected LDOS signal, the loop contribution and hence the HF, should be well observable.
	For the full LDOS plot, however, loop contribution is dominated by single contribution hence basically one can only observe single contribution in the full plot. Therefore, it is crucial to isolate the loop contribution in STM measurements. 
	%As the single impurity scattering always contributing concentric circles pattern, significantly different from the hyperbolic pattern from the loop contribution, the single contribution can be conveniently subtracted in experiments. For non-magnetic impurities, the dominant single-scattering contribution comes from the first-order term. For magnetic impurities, since electrons with spin up and spin down experience a scattering potential with opposite signs, single-scattering contributions from scattering processes of odd orders vanish. Thus, the dominant single-scattering contribution is from the second-order term, similar to the loop contribution.
	Here we proposed two different experimental proposals.
	
\begin{figure*}[tbp]
		\centering
		\includegraphics[width=0.75\textwidth]{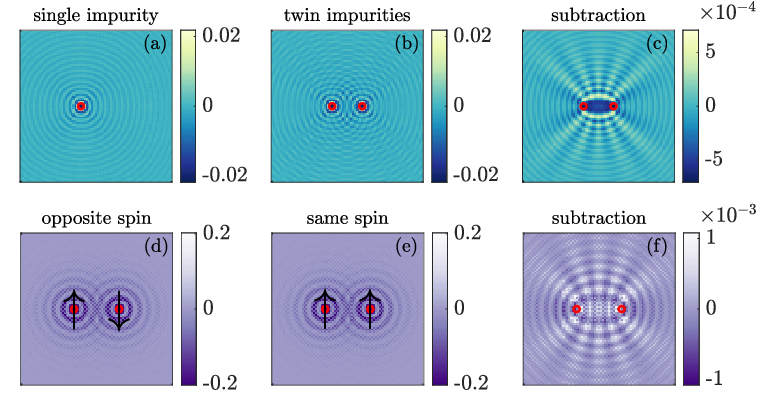}
		\caption{Proposals to isolate loop contributions in experiments. Generally, we can subtract the signal measured around single impurity (a) from the signal measured around twin impurities (b) to get a clear loop contribution signal (c). For magnetic impurities, we can perform the experiment two times, one time with the spins of two magnetic impurities pointing towards the opposite direction (d) and the other time with them pointing in same directions (e). Single contribution is greatly suppressed by subtracting the second signal from the first hence a clear loop signal (f) can be revealed. For (a), (b) and (c) we use a p+ip-wave state and non-magnetic impurities. For (d), (e) and (f) we take iron-based superconductor with sign-changing order parameter as an example.}
		\label{exp_mag}
	\end{figure*}

	Firstly, We propose a two-step measurement method to obtain the loop contribution from non-magnetic or magnetic impurities, which involves first measuring the interference pattern from a single impurity and performing another measurement for twin impurities (the second impurity can be introduced by STM atomic manipulation). The interference pattern from the second impurity alone can be obtained through translation of the data from the first impurity. By subtracting the single impurity signals from the interference pattern around twin impurities, we can isolate the loop contribution. Note that though single contribution cannot be completely cancelled due to multi-scattering processes (higher-order terms), loop contribution (second-order term) dominates in this method. This procedure is numerically demonstrated in Fig.~\ref{exp_mag}(a)-(c) and shows that the loop contribution from twin impurities can be readily extracted in experiments.
	
	Secondly, for magnetic impurities, an alternative technique entails performing two measurements for twin impurities with anti-aligned and aligned magnetic moments, as demonstrated in Fig.~\ref{exp_mag} (d), (e), and (f). The flip of magnetic moment may be achieved through the application of an external magnetic field or substitution of one impurity with another possessing an opposing magnetic moment using STM atomic manipulation. Note that, when scattering around magnetic impurities, since electrons with spin up and spin down experience a scattering potential with opposite signs, contributions from scattering processes of odd orders vanish. Also, single contribution to the second order of impurity potential cancels each other in these two configurations, while loop contribution to the second order remains. Therefore, in the subtraction plot, loop contribution dominates over single contribution. A clear hyperbolic fringe pattern indicating the loop contribution was obtained for iron-based superconductors featuring $s_{\pm}$-wave pairing, as depicted in Fig.~\ref{exp_mag} (d)-(f). The isolated loop contribution patterns surrounding the twin impurities can also differentiate between different pairing symmetries.

    \section*{IX. Patterns of the Fourier spectra}
    
    \begin{figure*}[tbp]
    		\centering
    		\includegraphics[width=0.75\textwidth]{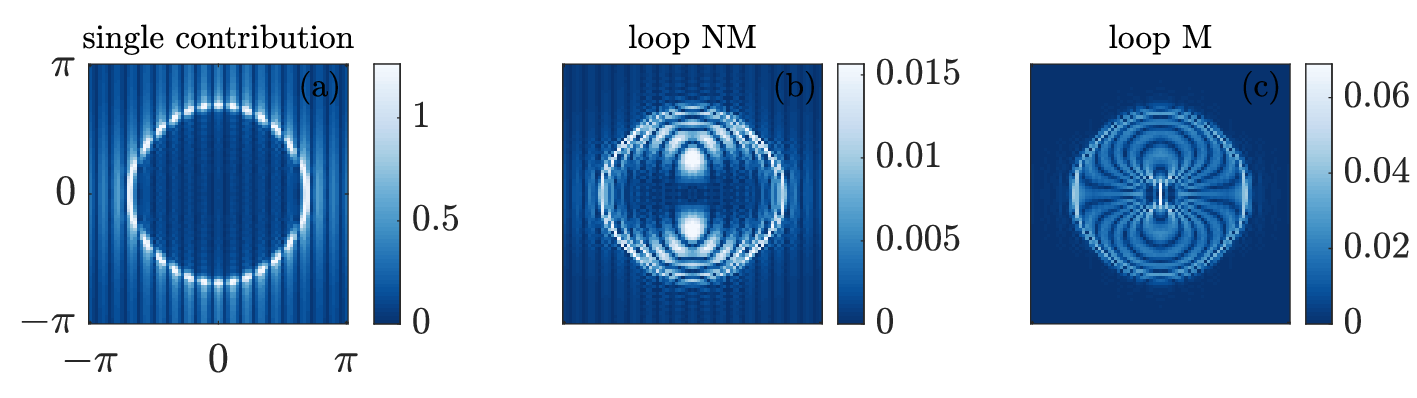}
    		\caption{Fourier spectra for twin impurity QPI for $s$-wave SC for (a) single contribution (b) loop contribution around NM impurities and (c) loop contribution around M impurities. $\mu = -3$,  $\Delta = 0.1$, $\omega = 1.1\Delta$, $d=18$.}
    		\label{s_FT}
    \end{figure*}
    
    \begin{figure*}[tbp]
    		\centering
    		\includegraphics[width=0.75\textwidth]{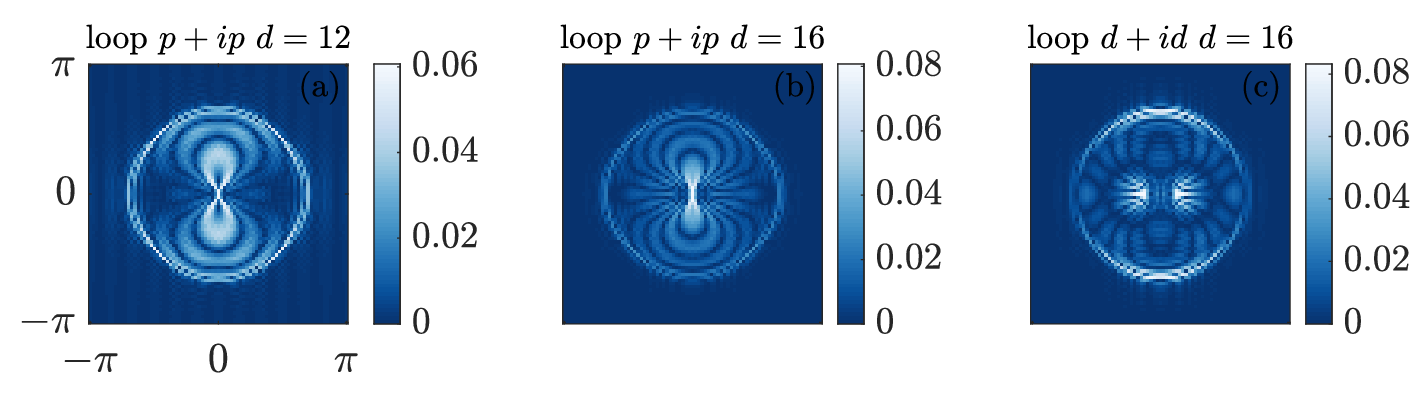}
    		\caption{Fourier spectra for the loop contribution in $p+ip$ ((a) $d = 12$ and (b) $d = 16$) and $d+id$-wave ((c) $d = 16$) SC. $\mu = -3$,  $\Delta = 0.01$, $\omega = 1.1\Delta$.}
    		\label{pd_FT}
    \end{figure*}

    In this section, we provide a detailed study of the Fourier transformations (FT) of the twin impurity QPI patterns.
    
    We begin with $s$-wave, for which we show the FT of the plots in Fig.~1 in the main text (see Fig.~\ref{s_FT}). For the single contribution, we see that the FT is basically just a ring of radius $2 k_F$, modulated by the interference pattern given by $|\cos(k_x \frac{d}{2})|$. The loop contribution also features a circle of radius $2 k_F$, indicating that spatial correlation with wavelength $k > 2 k_F$ is small. We find the existence of certain stripes within the circle, with frequency proportional to $\frac{1}{d}$ (the $d$ dependence of the strip is exhibited in the next figure), for both NM and M impurities. Their shapes, however, are very different. The FT of elliptical fringes (Fig.~\ref{s_FT} (b)) in real space also features elliptical dark curves, albeit the long and short axis are flipped. In contrast, the FT of hyperbolic fringes (Fig.~\ref{s_FT} (c)) exhibits drop-shape bright curves.

    We move on to show the FT for the loop contribution in $p+ip$ and $d+id$-wave SC (Fig.~\ref{pd_FT}). The role of distance $d$ in the FT plots is illustrated in the comparison of (a) and (b). We see that the shape of the stripes are the same, albeit the frequency of their appearance is proportional to $\frac{1}{d}$. The contrast between $p+ip$ and $d+id$ is manifest, which is essentially due to the difference of the phase structure of the pairing. Recall from Section.~\ref{analysis_pd} that for $p+ip$-wave, there is always a constant region between the two impurities. This is reflected in the peak around $k = 0$ in the Fourier spectrum. For $d+id$-wave, oscillations exist between the two impurities in real space, which is reflected in the two peaks on the $k_x$ axis in the Fourier spectra. Also, note that the overall pattern resembles a quadrupole, indicating the symmetry of the $d+id$-wave gap function.
    The significant difference in the Fourier spectra between the $p+ip$- and $d+id$-wave pairing provides an alternative way to distinguish them experimentally. 
    
    \subsection{Fourier spectra of the multi-band iron-based SC}
    
\begin{figure*}[tbp]
    		\centering
    		\includegraphics[width=0.75\textwidth]{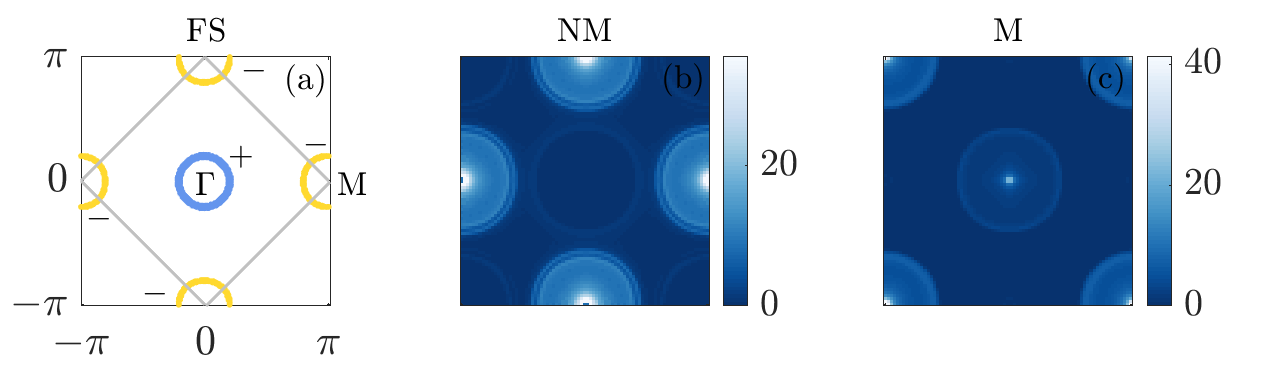}
    		\caption{Fourier spectra of single impurity QPI for iron-based superconductor for $s_{\pm}$-wave pairing (band structure and gap sign shown in (a)) around NM (b) and M (c) impurity. For (a) we only take into account one-times scattering process for simplicity. Parameters are the same as in Fig.~3 in the main text.}
    		\label{iron_single_FT}
    \end{figure*}

    The Fourier spectrum of the multi-band case is even more interesting. To facilitate understanding of the multi-band case, we firstly review the Fourier spectra of QPI around a single impurity. In Fig.~\ref{iron_single_FT} (b) and (c), we show the Fourier spectra of single impurity QPI for $s_{\pm}$-wave pairing iron-based superconductor around NM and M impurity, respectively. As analyzed in Ref.~\cite{Sykora}, due to the coherence factor of QPI, scattering off NM (M) impurity enhance sign-reversing (sign-preserving) momentum. Therefore, for NM impurity QPI peaks around $X$ and $Y$ points, while for M impurity QPI peaks around $\Gamma$ and $M$ points.
    
\begin{figure*}[tbp]
    		\centering
    		\includegraphics[width=0.75\textwidth]{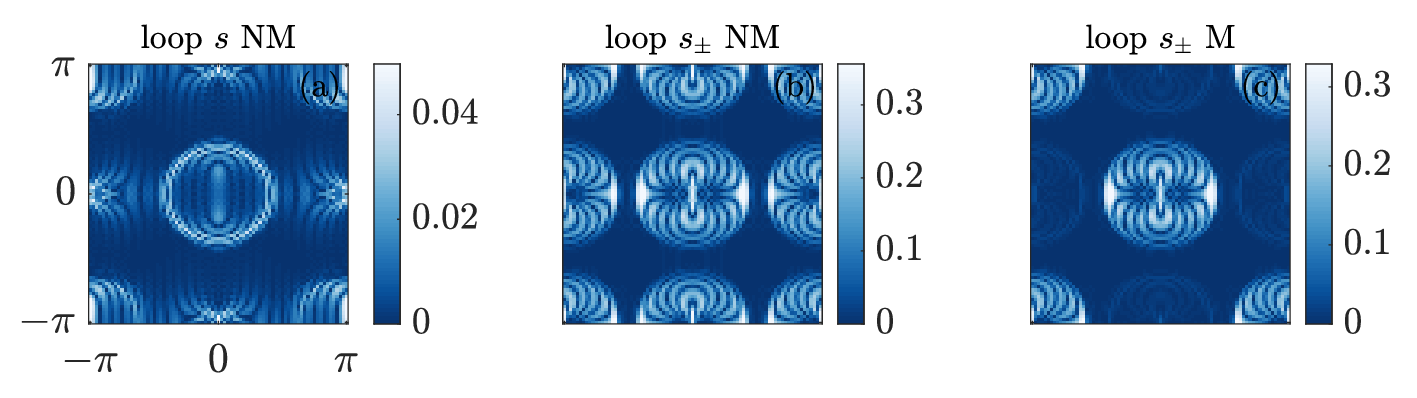}
    		\caption{Fourier spectra for the loop contribution for iron-based superconductor for (a) $s$-wave pairing, (b) $s_{\pm}$-wave pairing around NM impurities and (c) $s_{\pm}$-wave pairing around M impurities. Parameters are the same as in Fig.~3 in the main text.}
    		\label{iron_FT}
    \end{figure*}

    We then discuss the loop contribution with multiple scattering processes, shown in Fig.~\ref{iron_FT}.
    We find that the Fourier spectra show several with enhanced intensity and there are characteristic patterns inside each circle, in contrast to the single impurity scattering.
    For $s_{\pm}$-wave pairing, due to the emergence of HF as analyzed in the main text, there are drop-shape like stripes in each circle in both (b) for NM impurities and (c) for M impurities.
    The difference between them is due to the momentum selection rules discussed in the last paragraph. For (b), two-step scattering processes give rise to circles around $\Gamma$ and $M$ points, while three-step (and even higher order) scattering processes can give rise to circles around $X$ and $Y$ points.
    
    For NM impurities, the Fourier spectrum for $s$ and $s_{\pm}$-wave SC differs in two aspects. Firstly, the shape of the stripe in the central circle in (b) is different from (a) due to the emergence of hyperbolic fringes, the same reason as in Fig.~\ref{s_FT}. 
    Secondly, in (b) we see four circles with the same structure, centered at $\Gamma, X, Y, M$ points (of the unfolded Brillouin zone), while in (a) there is no circle around $X, Y$ points. This asymmetry is because this momentum involves scattering between electron and hole pockets. In (b) due the the sign-change of the gap on the electron and hole pockets, this asymmetry does not show up.
    %(Rigorous analysis of this asymmetry is too long and neglected here). 
    
    In summary, we emphasize that in the Fourier spectra of loop contribution in multi-band SC, both the location of the circles and the internal pattern inside each circle can provide crucial information about the properties of Fermi surfaces and sign change in the superconducting gaps.

\bibliography{supp}